\documentclass[12pt,preprint]{aastex}
\usepackage{emulateapj5,psfig}
\usepackage{epsfig}
\usepackage{amsmath}

\begin{document}
\title{Warm-Dense Molecular Gas in the ISM of Starbursts, LIRGs and ULIRGs}

\author{Desika Narayanan\altaffilmark{1}, Christopher
K. Walker\altaffilmark{1}, Christopher E. Groppi\altaffilmark{2,1}}
\altaffiltext{1}{Steward Observatory, University of Arizona, 933 N
Cherry Ave, Tucson, Az, 85721, 
\\E-mail: dnarayanan@as.arizona.edu, cwalker@as.arizona.edu} 
\altaffiltext{2}{National Radio Astronomy
Observatory, 949 N Cherry Ave, Tucson, Az., 85721. 
The National Radio Astronomy Observatory is a facility of the
 National Science Foundation operated under cooperative agreement by
 Associated Universities, Inc.
\\E-mail: cgroppi@nrao.edu} 



\begin{abstract}
 The role of star formation in luminous and ultraluminous infrared
galaxies (LIRGs, $L_{\rm IR} \geq$10$^{11} L_{\sun}$; ULIRGs $L_{\rm
IR} \geq$10$^{12} L_{\sun}$) is a hotly debated issue: while it is
clear that starbursts play a large role in powering the IR luminosity
in these galaxies, the relative importance of possible enshrouded AGNs
is unknown.  It is therefore important to better understand the role
of star forming gas in contributing to the infrared luminosity in
IR-bright galaxies. The J=3 level of $^{12}$CO lies 33K above ground
and has a critical density of $\sim$1.5$\times$10$^{4}$ cm$^{-3}$. The
$^{12}$CO(J=3-2) line serves as an effective tracer for warm-dense
molecular gas heated by active star formation.  Here we report on
$^{12}$CO (J=3-2) observations of 17 starburst spirals, LIRGs and
ULIRGs which we obtained with the Heinrich Hertz Submillimeter
Telescope on Mt. Graham, Arizona.

Our main results are the following: 1. We find a nearly linear
relation between the infrared luminosity and warm-dense molecular gas
such that the infrared luminosity increases as the warm-dense
molecular gas to the power 0.92; We interpret this to be roughly
consistent with the recent results of Gao \& Solomon (2004a,b).  2.
We find $L_{\rm IR}$/$M_{\rm H_2}$ ratios ranging from $\sim$10 to
$\sim$128 L$_\sun$/M$_\sun$ using a standard CO-H$_2$ conversion
factor of 3 $\times$ 10$^{20}$ cm$^{-2}$ (K km s$^{-1}$)$^{-1}$. If
this conversion factor is $\sim$an order of magnitude less, as
suggested in a recent statistical survey (Yao et al. 2003), then 2-3
of our objects may have significant contributions to the $L_{\rm IR}$
by dust-enshrouded AGNs.
\end{abstract}


\section{Introduction}

An important result from the $IRAS$ observatory was the discovery of
large numbers of galaxies which emit prodigiously in the infrared. The
$IRAS$ Bright Galaxy Survey ({\it BGS}, flux limited, $f_{60 \mu
m}\geq$5.4 Jy; Soifer et al 1987, 1989) reported on infrared fluxes of
330 galaxies.

An interesting subset of these objects, coined Luminous Infrared
Galaxies (LIRGs, $L_{\rm IR}\geq 10^{11} L_\sun$), and Ultraluminous
Infrared Galaxies (ULIRGs, $L_{\rm IR}\geq10^{12} L_\sun$), serve as
a unique tool to better understand the temporal evolution of starburst
phenomena (the field is reviewed by Sanders \& Mirabel, 1996). As
evidenced by morphological studies, ULIRGs are likely the product of
galaxy mergers, or strongly interacting galaxies (Sanders et al. 1988,
Lawrence et al. 1991, Leech et al. 1994). While many LIRGs are also
merger remnants, they can also be gas and/or dust-rich spirals.

The source of infrared luminosity in ULIRGs and high luminosity LIRGs
has been under contention for quite some time.  The luminous activity
from the central regions in LIRGs/ULIRGs suggests either a massive
starburst, dust enshrouded AGN, or some combination of the two. High
resolution ($\sim$1$\arcsec$) millimeter observations have shown large
amounts of molecular gas concentrated in the nuclear regions of these
galaxies (e.g. Scoville et al., 1989, Bryant \& Scoville 1999).
Numerical simulations have shown that galaxy mergers can efficiently
drive gas toward the nuclear regions of the remnants (Barnes \&
Hernquist 1992, Mihos \& Hernquist 1996). Nearly all bright $IRAS$
galaxies have been found to be rich in molecular gas (Tinney et
al. 1990, Sanders, Scoville \& Soifer 1991, hereafter SSS91). This
molecular gas is not only the nascent birthplace for massive star
formation, but may also act as fuel for a hidden AGN.  Additionally,
the processes may be linked. Several authors have suggested that the
large amounts of gas may lead to the buildup of large nuclear star
clusters which may augment the formation for a central AGN (Scoville
et al. 1989, Surace et al. 1998, Evans et al. 1999, Scoville et al. 2000).

Strong constraints must be placed on the interstellar medium (ISM) in
LIRGs/ULIRGs in order to better understand the source of luminosity
and their role in the evolutionary sequence of galaxies. Rotational
transitions in carbon monoxide are often used as a tracer for
molecular gas in the ISM.  The first extensive studies of the
molecular gas in LIRGs/ULIRGs via $^{12}$CO (hereafter, CO) line
emission were presented by Tinney et al. (1990), and SSS91. These
authors analyzed CO (J=1-0) emission from a sample galaxies with
$L_{\rm IR} \gtrsim$10$^{10}$ L$_{\sun}$.

Of particular interest in the ISM is the properties of warm, dense
gas, as it is this gas that directly traces star formation.  Because
CO (J=1-0) can be excited at relatively low temperatures ($\sim$5 K
above ground) and densities ($\sim$10$^3$cm$^{-3}$), it serves as a
good tracer for total molecular gas, but is relatively insensitive to
the warmer, denser gas directly involved in the star formation
process. In contrast, high lying rotational transitions of CO directly
trace warm, dense gas. The J=3 level of CO lies 33K above ground and
has a relatively high critical density of 1.5$\times$10$^{4}$
cm$^{-3}$.  The CO (J=3-2) transition can serve as a tracer of dense
molecular gas heated by active star formation, and thus as a
diagnostic for the starburst phenomena in these galaxies. Sensitivity
to the presence of dense gas is important in determining the
evolutionary state of LIRGs/ULIRGs. For example, in the well studied
galaxy, Arp 220, Solomon, Downes \& Radford (1992) presented single
dish data taken in CO (J=1-0), HCO+(J=1-0), and HCN (J=1-0). The CO
(J=1-0) spectra shows a single peak while the emission lines
from the high-density tracers HCO+ and HCN each showed two peaks.
Taniguchi \& Shioya (1998) discussed the double-horned profile seen in
the high-density tracers HCO+ and HCN as corresponding to the
starburst regions associated with two separate nuclei in this
galaxy. This model was further supported by high resolution
observations by Sakamoto et al. (1999).  It is therefore evident that
high-density gas tracers are vital to the understanding of molecular
gas in LIRGs and ULIRGs.

Several previous surveys of LIRGs and ULIRGs have relied on
millimeter-wave studies of HCN (J=1-0) in order to probe the
properties of the dense molecular component of these objects
(e.g. Solomon, Downes \& Radford, 1992, Gao \& Solomon 2004b, and
references therein). However, the low lying levels of HCN do not
necessarily trace gas that is both {\it warm and dense}.  For example,
the J=1-0 emission of HCN traces densities of $n$(H$_2$) $\gtrsim$ 3
$\times$ 10$^4$cm$^{-3}$; however, it only lies at a temperature of
$\sim$4.25K above the ground state.  Thus, while observations of this
transition will reveal the physical conditions of dense molecular gas,
both cool and warm, it does not necessarily probe the gas heated by
active star formation.  In order to study properties of the gas {\it
directly involved in star formation}, one must look toward lines that
have a high excitation temperature, as well as a high critical
density.

While there have been numerous studies of CO (J=1-0) emission
(Sanders \& Mirabel, 1996 and references therein), there are
relatively few studies of LIRGs and ULIRGs in higher lying CO
transitions.  Rigopoulou et al (1996) reported a CO (J=2-1) survey of
6 ULIRGs, and Yao et al. (2003) presented the first survey of CO(J=3-2)
emission of LIRGs and ULIRGs from the Scuba Local Universe Galaxy
Survey (SLUGS). Recent high-resolution studies of warm-dense molecular
gas in individual LIRGs have been performed by Iono et al. (2004) and 
Wang et al. (2004).

In order to better understand the role of star formation in LIRGs and
ULIRGs, and how it varies with infrared luminosity, we have conducted
a survey of CO (J=3-2) emission as a tracer for warm-dense molecular
gas in 17 normal starburst spirals, LIRGs and ULIRGs detected by the
{\it IRAS BGS} survey. We made these observations at the 10m Heinrich
Hertz Submillimeter Telescope (HHSMT) on Mt. Graham. A similar study
has been recently performed by Gao \& Solomon (2004a,b), using HCN
(J=1-0) as a tracer for dense molecular gas.  In \S2 we present the
observations and data; in \S3 we discuss the excitation conditions via
the CO (J=3-2)/CO(J=1-0) line ratio; in \S4 we interpret the line
profiles; in \S5 we discuss the data with respect to the dominant
source of infrared luminosity (starburst vs. AGN), and in \S6 we
summarize.

\section{Sample Selection and Observations}

The galaxies in this study were selected from the IRAS {\it BGS}
sample which was flux limited at $f_{60 \mu m} \geq$ 5.4 Jy. All but
two of the objects observed here have previously shown strong CO
(J=1-0) emission (Mirabel et al. 1990, Tinney et al. 1990, SSS91, Yao
et al. 2004) and were therefore known to be rich in molecular gas.
The galaxies were originally selected in their respective CO (J=1-0)
surveys from the {\it IRAS BGS} surveys, and in the case of Yao et
al. (2004), the SLUGS survey. The galaxies in these papers were chosen
for position on the sky, and thus not biased for any particular galaxy
properties. Our goal was to observe a number of galaxies over a wide
IR luminosity range: we observed 17 galaxies ranging in IR luminosity
from 10.41 $<$ log($L_{\rm IR}$) $<$ 12.39, spanning from starburst
spiral galaxies to ULIRGs. The infrared luminosities and 60$\mu$m
fluxes may be found in Table 1. All of the galaxies were chosen such
that they were not redshifted out of the atmospheric transmission
windows near 345 GHz. We additionally took care to observe objects
whose CO emitting region would be unresolved within the 23$\arcsec$
HHSMT beam (see also \S 3). We selected objects with a declination
greater than $\sim$-1$\arcdeg$. The full list of objects observed can
be found in Table 1.

We observed our sample of 17 galaxies on the HHSMT over three runs in
November 2003, March 2004 and January 2005.  We used the facility 345
GHz SIS receiver utilizing both polarizations.  The acousto-optical
spectrometer (AOS) was used as the backend, with a usable bandwith of
1GHz, and velocity resolution of 0.85 km s$^{-1}$.  For calibration
sources, we observed Orion, W3OH and IRC10216.  Telescope pointing was
checked every hour with observations of planets.  The observations
were made in beam switching mode with a throw of 60$\arcsec$, and a
chopping frequency of 4~Hz.  Typical integration time was $\sim$4
hours which yielded signal to noise ratios of about 10 in most
sources. Note, that due to time constraints, we were not able to
achieve this level of S/N for every object. Additionally, some of the
higher luminosity objects fell at a redshift such that their observed
frequencies were near atmospheric water lines, decreasing the S/N.
The weather conditions were good with a typical $\tau$ at 225 GHz of
$\sim$0.1.  We detected CO (J=3-2) emission in 15 out of the 17
galaxies in our sample\footnote{ Please note that while we did not
detect emission in NGC 3583, Yao et al. (2003) detected emission with
peak temperature of $\sim$0.1 K.  Baseline problems in our data
prevented us from ascertaining if we saw a line or simply baseline
ripple.}.


The data reduction was performed using the GILDAS CLASS package.  We
subtracted a linear baseline from the data, excluding points in the
emission line from the fit.  The data were then co-added, weighted by
the rms noise of each spectrum.  We smoothed the data to a resolution
between 3.2 km s$^{-1}$ and 25.6 km s$^{-1}$, depending on the noise
levels. We converted from an antenna temperature, $T_{\rm A}$, to main
beam temperature, $T_{\rm MB}$, by scaling by the main beam
efficiency, $\eta_{\rm mb}$, using $T_{\rm MB}$ = $T_{\rm A}/\eta_{\rm
mb}$.  Main beam efficiencies were measured to be $\sim$0.50 at the CO
(J=3-2) line frequency during the March 2004 observing run.  Observing
parameters may be found in Table 1.

While the galaxies were all resolved within our 23$\arcsec$ beam, the
nuclear gas and dust emitting regions were not.  In Figure 1, we show
the HHSMT beamprint overlaid on the optical images taken from the {\it
Digitized Sky Survey}\footnote{The Digitized Sky Surveys were produced
at the Space Telescope Science Institute under U.S. Government grant
NAG W-2166. The images of these surveys are based on photographic data
obtained using the Oschin Schmidt Telescope on Palomar Mountain and
the UK Schmidt Telescope. The plates were processed into the present
compressed digital form with the permission of these institutions.}
and the 2MASS images taken from the {\it 2-Micron All Sky Survey}. In
the same Figure we show the CO (J=3-2) spectrum we obtained for each
object in this study. The objects are arranged in order of increasing
$L_{\rm IR}$.

\section{CO (J=3-2)/(J=1-0) Line Ratio}
The ratio of intensities (K km s$^{-1}$) of the CO (J=3-2) line to the
CO (J=1-0) line, $R_{3}$, serves as a probe of the excitation
temperature and optical depth of the emitting gas.  A high ratio
($\gtrsim$1) indicates the gas is warm, and optically thin.

Because the CO (J=3-2) and (J=1-0) observations were taken at 
different telescopes with different beam sizes, certain considerations
must be taken into account when comparing the line intensities.  Both
the beam size, $b$, and angular extent of the sources on the sky, $s$,
play a role in determining the ratio:
\begin{equation}
R_3=\frac{I_{32}(s^2+b_{32}^2)}{I_{10}(s^2+b_{10}^2)}
\end{equation}
Where $I$ is the intensity (K km s$^{-1}$): 
\begin{equation}
I_{CO}=\int T_{\rm mb}(\rm CO){\it dV}
\end{equation}
In recent high resolution (2$\arcsec$-3$\arcsec$) millimeter-wave
observations of 7 LIRGs, Bryant \& Scoville (1999) find that nearly
all of the detected CO (J=1-0) emission is concentrated within the
central 1.6 kpc in 6 of the 7 objects.  Because it is unlikely that
there will be significant CO (J=3-2) emission where there is no CO
(J=1-0), we assume that all of the emitting gas in our sample is
confined within the same region, and, with the distances to each
object, compute the angular extent of emitting gas for each object.

The CO (J=1-0) data were taken at the IRAM 30m, Kitt Peak 12m, the
Nobeyama Radio Observatory, and the Swedish-European Submillimeter
Telescope (SEST) (Mirabel et al. 1990, Tinney et al. 1990, SSS91, Yao
et al. 2003).  The beam size of the 10m HHSMT at 345 GHz is
$\sim$23$\arcsec$.

The $R_3$ ratios are presented in Table 2. Our mean value of $R_3$ is
0.50. In a survey of CO (J=3-2) emission in 28 nearby galaxies,
Mauersberger et al. (1999) found a mean value of $R_3$ of 0.63.
Similarly, Yao et al. (2003) found a mean $R_3$ of 0.66 in their
survey of 60 LIRGs/ULIRGs.  The average $R_3$ for our
objects is larger than the average value seen in Galactic molecular
clouds of 0.4 (Sanders et al. 1993). The spread in the ratios in Table
2 suggest a variety of excitation conditions in the molecular gas in
our sample of galaxies. 


\section{Line Profiles}
The observed line profiles allow us to gain a better understanding of
the kinematics of the emitting molecular gas.  Krugel et al. (1990)
examined simple models of galaxies in order to explain their line
profiles.  They assumed the gas was distributed in a disk, and
constructed isothermal models of galaxies gridded in radius and
azimuth. They allowed the clouds to rotate around the center like a
rigid body out to a distance, $R_{\rm R}$, set as a free parameter,
and then forced the velocity to 200 km s$^{-1}$.  We use the results
of these models to illustrate possible kinematic signatures found in
our observed line profiles. The line profiles of the objects in our
sample tend to fall into three general categories: single Gaussian
(e.g. NGC 2276), double-peaked (e.g. Arp220), and three component
(e.g. NGC3079).  These can all be explained by models in which the
beam is larger than the emitting region, and a part of the flat
rotation curve is included.  While it is clear from Figure 1 that we
have resolved the optical emitting region of all of the galaxies in
this sample, as described in \S3, the bulk of the CO-emitting gas
likely remains unresolved within the 23$\arcsec$ beam.  Here, we
discuss the line profiles of the objects in this study, and how they
relate to other observations from the literature.

{\it \bf Single-Gaussian Profiles:} The single Gaussian line profile has a
width which varies based on viewing angle: if the galaxy is face on,
then the line width is dominated by the turbulent velocity among the
emitting clouds.  If, however, there is an inclination angle of
$i<$90$^{\circ}$, then the effects of the rotation of the galaxy are
evident in the line width. 

{\it NGC 7817}: The optical and 2MASS images (Figure 1) indicate our
  observations of the circumnuclear star-forming region are unresolved
  in this spiral galaxy. The CO (J=1-0) line profile appears to have a
  possible double-peak which would be consistent with the fact that we
  are viewing this galaxy slightly edge-on (SSS91). 

 {\it NGC 2276, NGC6701}: The narrow single-Gaussian line profiles for
  these galaxies are likely due to viewing these spirals $\sim$face
  on.

 {\it NGC 834}: The CO (J=1-0) profile in this spiral (Chini,
  Krugel \& Lemke 1996, Hattori et al. 2004) is difficult to interpret
  due to noise (SSS91, Young et al. 1995). It appears to be
  double-horned in CO (J=1-0). A large ring of star forming gas is
  seen in H$\alpha$ maps by Hattori et al. (2004): this ring may be
  the origin of the 1-0 double peak. This double peak may emerge in a
  CO (J=3-2) spectrum with a higher signal to noise ratio.

 {\it IRAS 23436+5257, IRAS 07251-0248}: The emission lines presented
  here for these galaxies are the first to be published. The objects
  appear to be disturbed and/or interacting galaxies (Figure 1).

 {\it UGC 5101}: The CO (J=3-2) spectrum presented here has an
  asymmetric profile with a high-velocity bulge. The HCN (J=1-0) line
  appears to be a more symmetric Gaussian (Gao \& Solomon, 2004b).
  The CO (J=1-0) spectrum, in contrast has a clear double-horned
  profile (Solomon et al. 1997).  Armus et al. (2004), discuss the
  presence of a buried AGN in this galaxy through the detection of [Ne
  V] 14.3 $\mu$m line. Farrah et al. (2003) describe this object as
  having an enshrouded AGN which plays a large role in contributing to
  the near-IR flux, but becoming more negligible at longer
  wavelengths. Many other authors have discussed the possibility of an
  enshrouded AGN in this object (e.g.  Imanishi, Dudley \& Maloney,
  2001, Imanishi et al. 2003, and references therein). Sanders et
  al. (1988) interpret the morphology as a late-stage merger.  NICMOS
  images show only a single nucleus (Scoville et al. 2000), making it
  unlikely that the possible double-peak is due to individual nuclei
  from progenitor galaxies in a merger. High resolution mapping will
  help to interpret this line profile better.

 {\it IRAS 17208-0014}: Gao \& Solomon (2004a) present HCN (J=1-0)
observations of this object, and Rigopoulou et al.(1996) present CO
(J=2-1) observations of this object. Our profile appears consistent
with both sets of data. The object has the optical spectrum of an HII 
region (Veilleux, Sanders \& Kim. 1999). Scoville et al. (2000) and
Farrah et al. (2003) classify IRAS 17208-0014 as being powered by a pure
starburst, with little to no AGN component. NICMOS images show this
galaxy to have a disturbed outer disk (Scoville et al. 2000).

{\it \bf Double Horned Profiles:} The double-horned profile was
evident in many of our objects as well.  There are two distinct
sub-categories: those with symmetric peaks (e.g. NGC 3094 and Arp
220), and those where one peak is higher than the other (e.g. NGC
992).  Krugel et al. model the symmetrical double-peaked objects as
galaxies that are unresolved, with the bulk of the emitting gas
confined to a circumnuclear ring.  However, as was discussed in \S1,
and below, there may be alternative explanations for NGC 828 and Arp
220.  The asymmetric double peak profile is more enigmatic.  Krugel et
al. suggest that if the gas is distributed symmetrically in the
galaxy's disk, such profiles might arise if the observation is
centered off the center of the disk.  Alternatively, this profile may
arise if the LIRG/ULIRG has two distinct nuclei from a merger, and the
emitting molecular gas from one has a larger velocity dispersion than
the other.

 {\it NGC 3094}: While our data for this object are sufficiently noisy
that a double peak is not clear, higher resolution data from Yao et
al. (2003) show the double peak to be evident. Imanishi (2000)
classifies this object has having a highly obscured AGN.

 {\it NGC 992}: While the CO (J=3-2) spectrum shows an asymmetric
  double-horned profile, the CO (J=1-0) profile is inconclusive.  The
  general shape is similar, but noise dominates the features enough to
  make it difficult to ascertain whether or not there are two
  individual peaks in the 1-0 profile (SSS91). The galaxy appears to
  be a spiral (Chini, Krugel \& Lemke 1996) undergoing a starburst
  (Ashby, Houck \& Matthews, 1995). It is likely, following the
  modeling of Krugel et al., that the asymmetric line profile arises
  from a telescope pointing offset from the center of the CO disk.

 {\it NGC 828}: Hattori et al. (2004) observed H$\alpha$ in this
  object, and describe it as a disturbed spiral galaxy.  Kinematic
  evidence from high-resolution CO (J=1-0) mapping suggest that this
  galaxy is an ongoing merger (Wang, Scoville \& Sanders, 1991).  CO
  (J=1-0) spectra (SSS91, Young et al. 1995) appear to display two
  peaks, although the double-horn is significantly more asymmetric
  than is seen in the two major peaks in the 3-2 spectra presented in
  this study.

 {\it NGC 7771}: The observation may be slightly off from the
  optical center ({\it DSS} images, Figure 1), although the entire gas
  emitting region appears to remain unresolved within the beam ({\it
  2MASS} images, Figure 1). The CO (J=1-0) profile looks
  single-Gaussian. The double peak is likely due to the rotational
  structure of the galaxy.

 {\it Arp 220}: This object is well studied in the literature, and we
  will only briefly discuss the line profile. As discussed in \S1, the
  CO (J=1-0) spectrum of Arp 220 shows a single Gaussian peak, while
  the spectrum from high-density tracers (e.g. HCN J=1-0, CO J=3-2)
  reveal a double peak.  The double peaks are interpreted to belong to
  each nuclei of the progenitor galaxies (Taniguchi \& Shioya, 1998).

{\it \bf Triple-Peaked Profiles:} Of our sample, NGC 3079 and NGC
6286 were best fit by three Gaussians; When fit to a single Gaussian,
the FWHM's of these objects are $\sim$450 km s$^{-1}$.  Krugel et
al. observed a similar feature in the starburst galaxy, Mrk 1034, and
modeled this as the spectral signature of a resolved galaxy where the
inclination was such that both the central core, and extended gas with
a flat rotation curve were being observed.  This type of model
produces the symmetric spectra seen in some of our observations, with
the height of the central peak rising with respect to the two outer
peaks as the inclination angle of the galaxy drops.

{\it NGC 3079}: This galaxy appears to fit the Krugel et
  al. model. Indeed the 2MASS image shows that the beam extends
  significantly beyond the nuclear dust and gas emitting
  region. Tinney et al. (1990) presented a CO (J=1-0) spectrum of NGC
  3079 compiled from a map; the profile appears to be an asymmetric
  double-horn.

 {\it NGC 6286}: The optical and 2MASS images reveal this
  observation to be of an edge on spiral. The asymmetric line profile
  was best fit by three Gaussians, although it is likely to simply be
  a double-horned spectrum that was observed off of the CO emission
  center.  Indeed, the CO (J=1-0) line profile appears more symmetric, 
  although the presence of more than one peak is difficult to determine 
  given the noise (SSS91).

\section{Source of IR Luminosity}



The search for the driving source of high luminosity LIRGs and ULIRGs
has been the subject of numerous observational and theoretical
studies.  While it is clear that there is active star formation in
these objects, there remains evidence that dust-enshrouded active
nuclei may play an important role as well (e.g. Yun \& Scoville,
1998). In this section, we investigate the source of IR luminosity in 
the high-luminosity LIRGs and ULIRGs in our sample.

\subsection{$L_{\rm IR}$ versus $L_{\rm CO \ J=3-2}$}

Correlations between infrared luminosity, and the luminosity due to
high density tracers such as HCN have been discussed extensively
(Solomon et al. 1992, Gao \& Solomon 2004a,b).  In a recent survey of
normal (spiral) galaxies, LIRGs and ULIRGs, Gao \& Solomon (2004b)
found a linear relationship between $L_{\rm IR}$ and $L_{\rm HCN}$
over three orders of magnitude in infrared luminosity; this suggests
that over the IR luminosity range spanned (log($L_{\rm IR}$)
$\lesssim$ 12.36) the dense molecular gas (and thus, high mass star
formation) is the dominant form of IR emission in both the LIRGs and
ULIRGs.  Additionally, Carilli et al. (2004) have shown for high
redshift ULIRGs that the correlation between IR luminosity and HCN
luminosity is nearly linear as well. These correlations suggest that
at some level, the dense interstellar medium and infrared luminosity
in these objects are intertwined.  Since CO (J=3-2) is an indicator of
warm-dense gas, we use our dataset combined with those of other works
to further test this hypothesis.  Similar to the Gao \& Solomon study,
we probe infrared luminosities up to log($L_{\rm IR}$ )=12.39.

  We obtained values for $L_{\rm IR}$ from SSS91 and the {\it IRAS
Revised Bright Galaxy Survey} (Sanders et al. 2003).  We calculated
$L_{\rm CO}$ using
\begin{equation}
{L_{CO}}\approx\pi/(4 \rm ln 2)\theta_{\rm mb}^2{\it I}_{\rm 32} \
{\it d}_L^2(1+{\it z})^{-3}
\end{equation} 
where $\theta_{\rm mb}$ is the FWHM of the telescope Gaussian beam,
 and $d_{\rm L}$ is the luminosity distance.  The data for $L_{\rm
 CO}$ are presented in Table 2. So that we may increase the number of
 objects in our analysis, we included 14 objects which we detected
 (please see caption of Figure 2 concerning exclusions), as well as 40
 objects\footnote{All the objects for which Sanders et al. 2003
 reported an infrared luminosity} studied by Yao et al. (2003). In
 order to account for calibration differences in the data sets, we
 used Arp 220 as a common calibrator, and scaled our $L_{\rm CO}$
 values such that the Arp 220 CO luminosities matched.  This scaling
 factor was 0.26.

We present $L_{\rm IR}$ vs. $L_{\rm CO \ J=3-2}$ in Figure 2.  We fit the data 
using a linear $\chi^2$ minimization routine, and recovered for the fit:
\begin{equation}
{\rm log}(L_{\rm IR})=0.92 (\pm 0.07)\times {\rm log}(L_{\rm CO \
J=3-2})+3.28(\pm 0.60)
\end{equation}
This result is roughly consistent with the linear slope found by Gao \& 
Solomon (2004a,b).

If the observed FIR emission is produced primarily by heating from
massive stars, the star formation rate (SFR) can be given as a
function of $L_{\rm IR}$ (Kennicutt, 1998):
\begin{equation}
{\rm SFR}(M_\sun \ {\rm yr}^{-1}) \approx 2 \times 10^{-10}(L_{\rm IR}/L_\sun)
\end{equation}
 Previous studies of the relationship between $L_{\rm IR}$ and L$_{\rm
CO \ J=1-0}$ have found a nonlinear relationship such that the
infrared luminosity increases as the CO (J=1-0) luminosity to the
$\sim$1.25. Because $L_{\rm CO \ J=1-0}$ is proportional to the total
molecular gas mass (Appendix A of SSS91), this has been interpreted as
an increase in star formation efficiency (SFE), defined as the SFR
over the total molecular gas mass, as a function of SFR (e.g. Solomon
et al. 1992).  $L_{\rm IR}$/CO(J=1-0) serves as a tracer for star
formation efficiency as a function of all available molecular
gas. $L_{\rm IR}$/HCN(J=1-0) serves as a tracer for SFE as a function
of all available {\it dense, star-forming gas} (e.g. Gao \& Solomon,
2004b). In contrast, CO(J=3-2) traces warm, dense gas heated by
embedded stars (\S1) and thus $L_{\rm IR}$/CO(J=3-2) probes the
infrared luminosity as a function of {\it recent star formation}. It
follows then, that the linear relationship seen between $L_{\rm IR}$
and $L_{\rm HCN}$ implies that the SFR per unit {\it dense} molecular
gas mass remains constant through log($L_{\rm IR}) \lesssim $ 12.3;
this result is confirmed by our observed relationship between $L_{\rm
IR}$ and $L_{\rm CO \ J=3-2}$ which shows a nearly constant SFR as a
function of warm-dense molecular gas heated by active star formation.
These observations strengthen the conclusion of Gao \& Solomon (2004b)
that the IR luminosity in LIRGs and ULIRGs up to an IR luminosity of
at least log($L_{\rm IR}$) $\sim$ 12.3 is primarily driven by heating
from O and B stars.


  The observed linear relationships suggest that star formation does
play an important role in powering the infrared luminosity in LIRGs
and ULIRGs.  However, it may be that this relationship breaks down at
higher IR luminosities as possible AGN contribution becomes more
important.  As an example, we extrapolate our fit to observations of
Hyperluminous Infrared Galaxies (HLIRGs). Extrapolating from our fit,
a galaxy with log$(L_{\rm IR})$$\sim$13 would require $\sim$4
$\times$10$^{11}$M$_{\sun}$ of warm-dense molecular gas in the nuclear
regions. For comparison, the HLIRG FSC 10214+4274 has $L_{\rm
FIR}$=1.8$\times$10$^{14}$L$_{\sun}$ and CO (J=3-2)-traced warm-dense
gas mass of $M_{\rm H2}$= 2.2$\times$10$^{11}$ M$_{\sun}$ (an upper
limit as the luminosity may have been enhanced by gravitational
lensing, Close et al. (1995)); similarly, HLIRG FSC 15307+3252 has
$L_{\rm FIR}$=1.3$\times$10$^{13}$L$_{\sun}$ and has an upper limit of
warm-dense gas mass (as measured from CO J=4-3 observations) of
$M_{\rm H2} \lesssim$ 10$^{10}$ M$_{\sun}$.  Both are believed to have
central AGN as their main source for infrared luminosity (Yun \&
Scoville 1998).

The disparity in the measured versus predicted amount of
warm-dense gas in these HLIRGs suggests that at higher infrared
luminosities, there may exist a strong deviation from the fit
described by Equation 4 which may be evident from the higher
luminosity points in Figure 2.  This implies that the infrared
emission may grow non-linearly with increasing emission from
warm-dense molecular gas.

 It may be that there is an interplay between an AGN and star
formation contribution to $L_{\rm IR}$. The case has been mounting for
an evolutionary sequence between hierarchical mergers and AGN
formation (e.g. Scoville, 2003).  It has recently become clear that
most galaxies contain supermassive black holes (Kormendy \& Richstone,
1995, and references therein).  There is also evidence that there is a
connection between black hole growth and central ISM physics in the
remnants of galaxy mergers.  Recent SPH simulations have shown that
AGN feedback in galaxy merger remnants can rapidly quench star
formation after the initial merger-induced starburst (Springel, Di
Matteo \& Hernquist 2004a,b). Starbursts may be the dominant source of
IR luminosity in less IR-luminous galaxies, while the AGN contribution
becomes more important in higher luminosity objects. Quenching of star
formation owing to AGN feedback may be responsible for a steepening of
the slope in the $L_{\rm IR}$ versus $L_{\rm CO }$ plot for the
galaxies with log($L_{\rm IR})$ $\gtrsim$ 12.3 L$_{\sun}$.  Indeed, some
35\%-50\% of ULIRGs with $L_{\rm IR}$ $\geq$ 10$^{12.3} L_{\sun}$ show
AGN activity from optical and NIR spectra (Tran et al. 2001; Veilleux,
Kim \& Sanders, 2002). This luminosity is approximately where our
sample of LIRGs and ULIRGs ends, and a location for a possible
deviation from the fit in Figure 2.  Additionally, Farrah et
al. (2002) discuss coeval starburst and AGN activity in a sample of
SCUBA HLIRG sources.
 
More CO (J=3-2) observations of ULIRGs with $L_{\rm
IR}\gtrsim10^{12.3}$L$_{\sun}$ are needed to test these
hypotheses.  Theoretical models testing these concepts will be
presented in an upcoming paper.

\subsection{$L_{\rm IR}$ versus  Mass of Star-Forming Molecular Gas}

One of the main arguments for the existence of a dust enshrouded AGN
as the main power source in luminous infrared galaxies is a SFE
($L_{\rm IR}$/M$_{\rm H_2}$) larger than that seen in normal spiral
galaxies.  If the sole source of energy is a nuclear starburst, then,
assuming a reasonable initial mass function, there should be an
Eddington-like limit on the star formation efficiency given by
SFE $\sim$500 $L_{\sun}$/$M_{\sun}$: objects with ratios significantly
higher than this are likely to require an additional source of energy
(Scoville, Yun \& Bryant 1997; Scoville, 2003).

The gas mass is obtained using a linear relation between the CO
luminosity and $M_{\rm H2}$ (Young \& Scoville, 1982; Tinney et al.,
1990; SSS91).
\begin{equation}
M_{H2}=4.82L_{CO}(M_{\sun})
\end{equation}
where \rm $L_{\rm CO}$ is in K km s$^{-1}$.  This relation is derived
 assuming a CO-H$_2$ conversion factor of 3 x 10$^{20}$ cm$^{-2}$ (K km
 s$^{-1}$)$^{-1}$, and was derived for Galactic GMCs (Scoville \&
 Sanders 1987). However, evidence exists that implies this conversion
 factor may not hold for external galaxies (Maloney 1989).  Through a
 Large Velocity Gradient (LVG) analysis, Yao et al. (2003) suggested
 that for LIRGs/ULIRGs, the CO-H$_2$ conversion factor lies between 1.3 x
 10$^{19}$ and 6.7 x 10$^{19}$ cm$^{-2}$ (K km s$^{-1}$)$^{-1}$, or a
 factor of 0.04 to 0.22 times the conventional conversion factor.
 Nevertheless, we adopt the Galactic value in order to facilitate
 comparisons with published analyses of CO (J=1-0) observations.

While the ratio of infrared luminosity to total gas mass may be a
strong indicator as to the source of infrared luminosity, it is the
{\it dense} gas mass that is expected to more directly trace star
formation.  Thus, a prudent test is to examine the ratio of $L_{\rm
IR}$ to $M_{\rm H2, dense}$.  Gao \& Solomon (2004a,b), found that
this ratio remains between $\sim$20 and $\sim$200, independent of
total IR luminosity for normal spirals, LIRGs and ULIRGs.  They
additionally reported that the average $L_{\rm IR}$ to $M_{\rm H2,
dense}$ in their sample was $\sim$90L$_{\sun}$/$M_{\sun}$, comparable
to molecular cloud cores (Mooney \& Solomon 1988).  These ratios were
found through analysis of HCN (J=1-0).

We present the ratio $L_{\rm IR}$/$M_{\rm H2, warm-dense}$ for our
sample of LIRGs in Table 2.  As is evident, none of the objects have a
ratio $L_{\rm IR}$/$M_{\rm H2}>$500$L_{\sun}$/$M_{\sun}$.  There are a
few important caveats to note. This number could be significantly
skewed if our mass estimates are off.  Certainly (following the
derivation presented in Appendix A of SSS91) if the CO-to-H$_2$
conversion factor is, as derived by Yao et al. (2003), an order of
magnitude lower than the conventional values, the $L_{\rm IR}$/M$_{\rm
H2}$ values would increase accordingly.  Using the CO-H$_2$ conversion
factors calculated by Yao et al., we find 2-3 of our objects have
ratios higher than 500, indicating the possibility of an AGN component
to the observed far infrared luminosity.  It should also be noted that
the ratios presented here are of star formation rate per unit {\it
warm, dense} gas mass - this hints at the possibility that the actual
{\it star forming} gas available is not enough to produce the infrared
luminosities seen in some objects. Additionally, it is not clear that
the CO-to-H$_2$ conversion factor is the same in all galaxies.  Better
constraints on variations in the CO-to-H$_2$ conversion factor are
necessary in order to use this method to make a robust statement
concerning the source of IR luminosity.




\section{Summary}
We have presented single-dish CO (J=3-2) observations of a sample of
starbursts, LIRGs and ULIRGs in order to study properties of the
warm-dense, star forming gas.  We detected emission in 15 out of 17
galaxies within our noise limits. We have found a nearly-linear
relationship between the infrared luminosity and amount of warm-dense
gas, confirming the recent results by Gao \& Solomon for galaxies with
log($L_{\rm IR}$)$\lesssim$12.3.  Depending on the value of the
adopted CO-H$_2$ conversion factor, our analysis suggests that star
formation alone may not be sufficient to generate the IR emission
observed in the more luminous galaxies.

\acknowledgements

We would like to thank the entire HHSMT staff for continual
 support. We additionally acknowledge R.S. Bussmann, Abigail Hedden,
 and Elaina Hyde for their assistance during observations.  We thank
 the referee for helpful comments that have greatly improved the
 quality of this manuscript, and for a prompt review. D.N. would like
 to thank Craig Kulesa, and Bob Stupak for helpful conversations, and
 acknowledges the NSF Graduate Research Fellowship Program for funding
 during part of this study.

The HHSMT is operated by the Arizona Radio Observatory (ARO), Steward
 Observatory, University of Arizona, with partial funding from the
 Research Corporation. This research has made use of the NASA/IPAC
 Extragalactic Database (NED) which is operated by the Jet Propulsion
 Laboratory, California Institute of Technology, under contract with
 the National Aeronautics and Space Administration. This research has
 made use of the NASA/ IPAC Infrared Science Archive, which is
 operated by the Jet Propulsion Laboratory, California Institute of
 Technology, under contract with the National Aeronautics and Space
 Administration

\clearpage

\begin{deluxetable}{cccccccccc}
\tabletypesize{\scriptsize} 
\tablecaption{Observation Information\tablenotemark{1}}
\tablewidth{0pt} 
\tablehead{ \colhead{Object} & \colhead{RA (J2000)}
&\colhead{Dec (J2000)} &\colhead{cz} &\colhead{Distance}\tablenotemark{2}
&\colhead{Date Observed} & Reference\tablenotemark{3}
&\colhead {log$L_{\rm IR}$}\tablenotemark{4}
&\colhead{$f_{60 \mu m}$}\tablenotemark{4}
&\colhead{Detection?}}
\startdata

        &             &              & km s$^{-1}$& Mpc& &  &$L_\sun$ & Jy  &    \\
\hline
NGC 7817& 0:04:10.1   & 20:46:30.6   & 2315 & 31.9 & 11/03  & 1&10.41 &* &Yes\\
NGC 3583& 11:14:28.0  & 48:17:38.5   & 2168 & 35.09 & 11/03 & 3&10.54 & 7.43  &No\\
NGC 3079& 10:02:16.3  & 55:39:46.9   & 1142 & 18.19 & 03/04 & 3&10.73 & 50.67 &Yes\\
NGC 3094& 10:01:38.3  & 15:45:10.3   & 2404 & 37.04 & 11/03 & 3&10.73 & 10.88 &Yes\\
NGC 2276& 7:28:27.8   & 85:45:09.8   & 2380 & 35.83 & 03/04 & 2&10.81 & 14.29 &Yes\\
NGC 834 &  2:11:14.4  &  37:41:11.2  & 4687 & 61.48 & 03/04 & 1&10.94 & 6.65  &Yes\\
NGC 992 & 02:37:25.4  &  21:05:47    & 4136 & 53.83 & 03/04 & 1&11.02 & 11.40 &Yes\\
NGC 6701& 18:43:14.9  & 60:38:58.1   & 3933 & 56.64 & 03/04 & 1&11.05 & 10.05 &Yes\\
NGC 6621& 18:12:54.9  & 68:21:27.0   & 6234 & 86.42 & 03/04 & 1&11.23 & 6.78  &No\\
NGC 828 & 02:10:09.42 &  39:11:26.29 & 5371 & 70.37 & 11/03 & 1&11.31 & 11.46 &Yes\\
NGC 6286& 16:58:36.6  & 58:55:25.7   & 5637 & 79.78 & 11/03 & 1&11.32 & 9.24  &Yes\\
NGC 7771& 23:51:36.2  & 20:08:00.4   & 4336 & 57.11 & 11/03 & 1&11.34 & 19.67 &Yes\\
IR23436+5257 & 23:46:05.8 & 53:14:00 & 10233 & 134.78 & 01/05 &* &11.51 & 5.66 & Yes\\
UGC 5101     &09:32:10.5 & 61:21:22  & 11785 & 158.61 & 01/05 & 5& 11.95& 11.68 & Yes\\
Arp 220 & 15:35:05.2  & 23:29:22.7   & 5450 & 79.90 & 11/03 & 3&12.21 & 104.09&Yes\\
IR07251-0248 & 07:27:37.5 & -02:54:55& 26257 & 343.74 & 01/05 &* &12.32 & 6.49 & Yes\\
IR17208-0014 & 17:23:21.4 & -00:17:00& 12852 & 175.68 & 01/05 &4 & 12.39& 32.13& Yes\\

\enddata
\tablenotetext{1}{All objects are in order of increasing $L_{\rm IR}$.}
\tablenotetext{2}{Distances given are proper distances, and are taken from
Sanders et al. 2003 except for NGC 7817 which is from SSS91. 
Values for cz taken from the same references.}

\tablenotetext{3}{References are for J=1-0 data: 1: Sanders et
al. 1991, 2: Tinney et al. 1990, 3: Yao et al. 2003, 4: Mirabel et
al. 1990,  5: Gao \& Solomon, 2004b. An asterisk denotes no CO (J=1-0) observations that we could find.}
\tablenotetext{4}{$L_{\rm IR}$ and 60$\mu$m flux from Sanders et al. 2003,
except for NGC 7817 which was not given.}

\end{deluxetable}

\clearpage

\begin{deluxetable}{ccccccccc}
\tabletypesize{\tiny} 
\tablecaption{Physical Properties}
\tablewidth{0pt} 
\tablehead{ \colhead{Object} & \colhead{Intensity}
&\colhead{$\pm$} &\colhead{$R_3$}\tablenotemark{1} &\colhead{L$_{CO}$} &\colhead{$\pm$}
&\colhead {M$_{\rm H2}$ CO(3-2)} &\colhead{$\pm$}
 &\colhead{$L_{\rm IR}$/M$_{\rm H2 \ warm,dense}$}
}\startdata
 & K km s$^{-1}$ & K km s$^{-1}$ & &
10$^9$K km s$^{-1}$ pc$^2$ & 10$^9$K km s$^{-1}$ pc$^2$ &10$^9$M$_{\sun}$ &10$^9$M$_{\sun}$ & L$_{\sun}$/M$_{\sun}$\\
\hline 
NGC 7817& 2.94         & 0.07    & 0.03  & 0.04    &   0.001  &  0.20  & 0.005& 128.52\\
NGC 3079& 183.46       & 3.18    & 0.51  & 0.852   &   0.01   &  4.11  & 0.07 & 13.07\\
NGC 3094& 15.10        & 0.41    & 0.60  & 0.290   &   0.008  &  1.40  & 0.04 & 38.69\\
NGC 2276& 12.23        & 0.20    & 0.12  & 0.219   &   0.004  &  1.06  & 0.02 & 60.91\\
NGC 834 & 11.15        & 0.37    & 0.28  & 0.585   &   0.02   &  2.82  & 0.09 & 30.89\\
NGC 992 & 21.16        & 1.25    & 0.34  & 0.852   &   0.05   &  4.11  & 0.24 & 25.48\\
NGC 6701& 28.80        & 0.42    & 0.45  & 1.28    &   0.02   &  6.19  & 0.09 & 18.13\\
NGC 828 & 59.21        & 1.77    & 0.50  & 4.06    &   0.121  &  19.56 & 0.58 & 10.44 \\
NGC 6286& 42.61        & 0.13    & 0.65  & 3.75    &   0.01   &  18.08 & 0.06 & 11.56\\
NGC 7771& 19.28        & 0.22    & 0.32  & 0.873    &   0.01   &  4.2 & 0.05 & 52.07\\
IR23436+5257 & 20.05  & 1.19     &    *   & 4.96    &   0.29   &  23.9  & 1.4  & 13.54 \\
UGC 5101 & 7.17       & 1.1      &  0.50 & 2.45  &   0.38   &  11.8  & 1.8  &75.53\\
Arp 220 & 89.96        & 1.91    & 1.60  & 7.95    &   0.17   &  38.30 & 0.81 & 42.34\\
IR07251-0248 & 7.70   & 0.31     &    *   & 11.79   &   0.47   &  56.8  & 2.3  & 36.78 \\
IR17208-0014 & 10.49  & 1.11     &    0.42 & 4.37    &   0.46   &  21.1  & 2.2  & 116.33 \\

\enddata
\tablenotetext{1}{Asterisk denotes no CO (J=1-0) data available.}

\end{deluxetable}

\clearpage



\figcaption[f1.ps]{ Each 3-panel row contains the following
information from left to right: The {\it DSS} image with the
23{$\arcsec$} HHSMT beam overlaid; The {\it 2MASS} image with the
23{$\arcsec$} HHSMT beam overlaid; The CO (J=3-2) emission line
obtained at the HHSMT in this study.  }

\figcaption[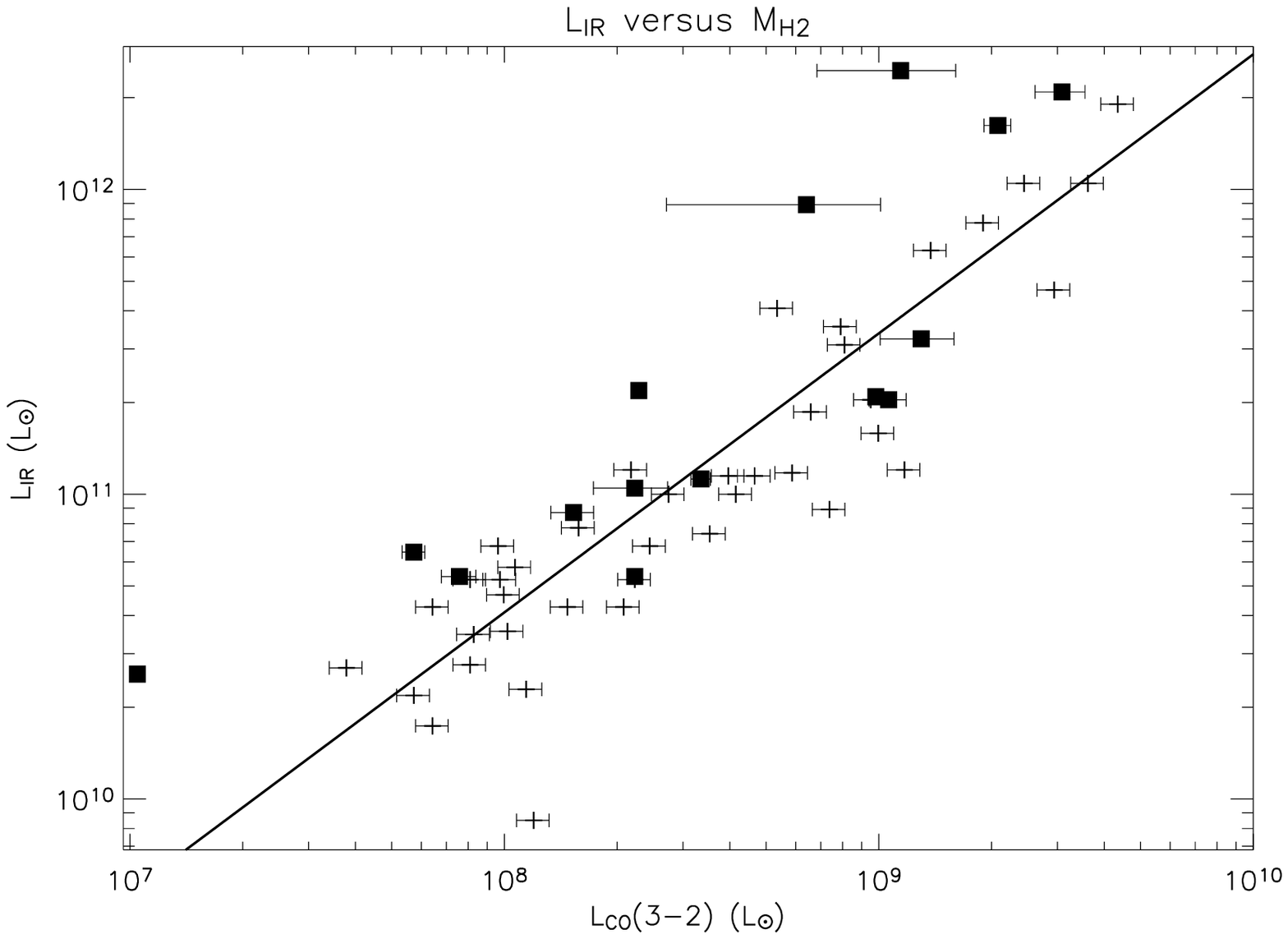]{ $L_{\rm IR}$ vs. L$_{\rm CO \ 3-2}$. The slope of
the fit is 0.92. Crosses are data points from Yao et al. (2003) while
filled in squares are data taken in this study.  Errors for objects
from our sample derived from errors in intensity calculations.  We
estimate an error of 10\% for the objects taken from the Yao et
al. (2003) sample. No error bars are given for the $L_{\rm IR}$ data
points.  We have excluded NGC7817 from the fit (but left its point on
the plot) as it has an extremely small $L_{\rm CO}$ with respect to
the locus of objects we observed.  When including this object, the
slope changes to 0.85$\pm 0.07$}

\newpage
\plotone{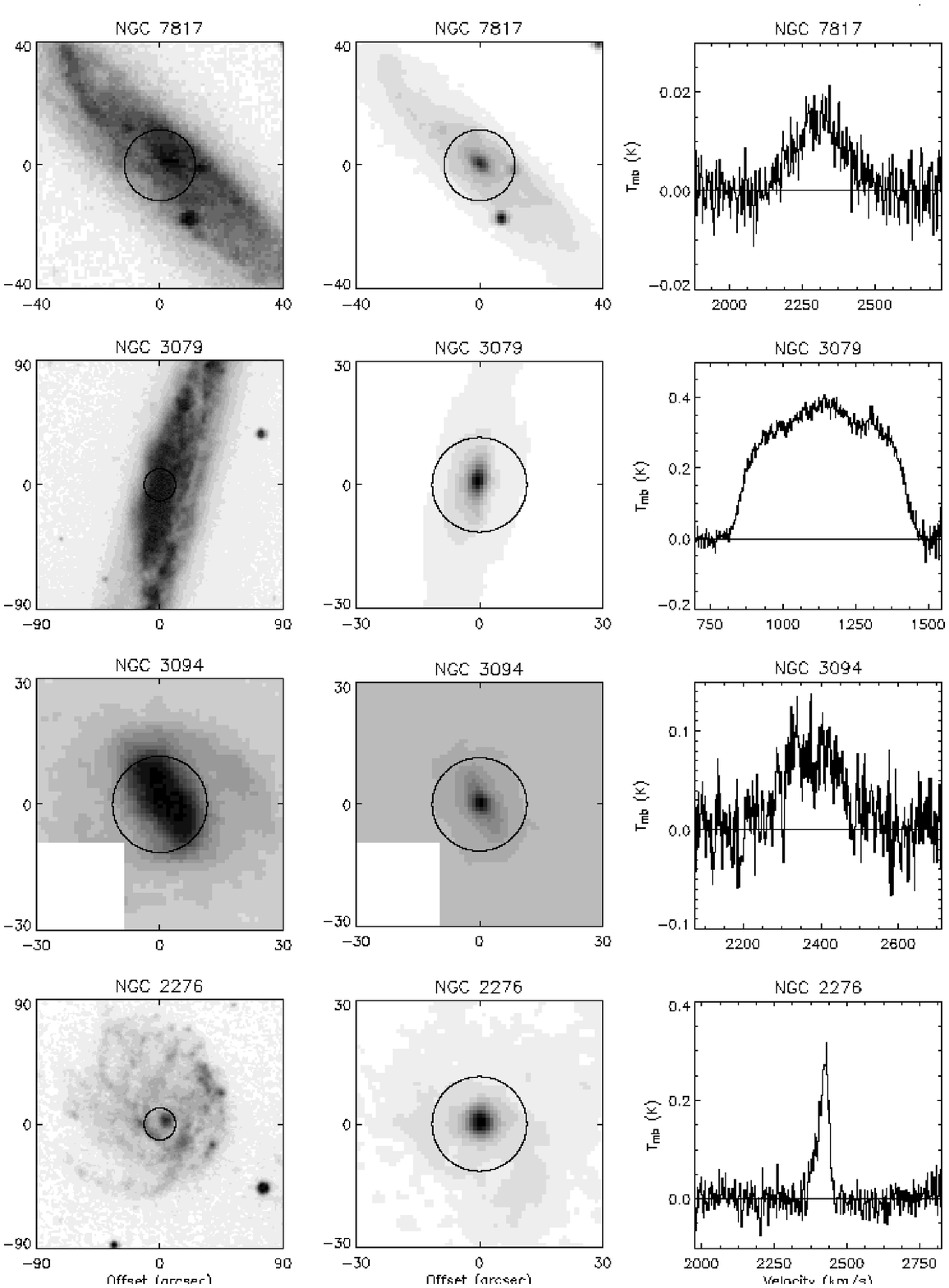}
\newpage
\plotone{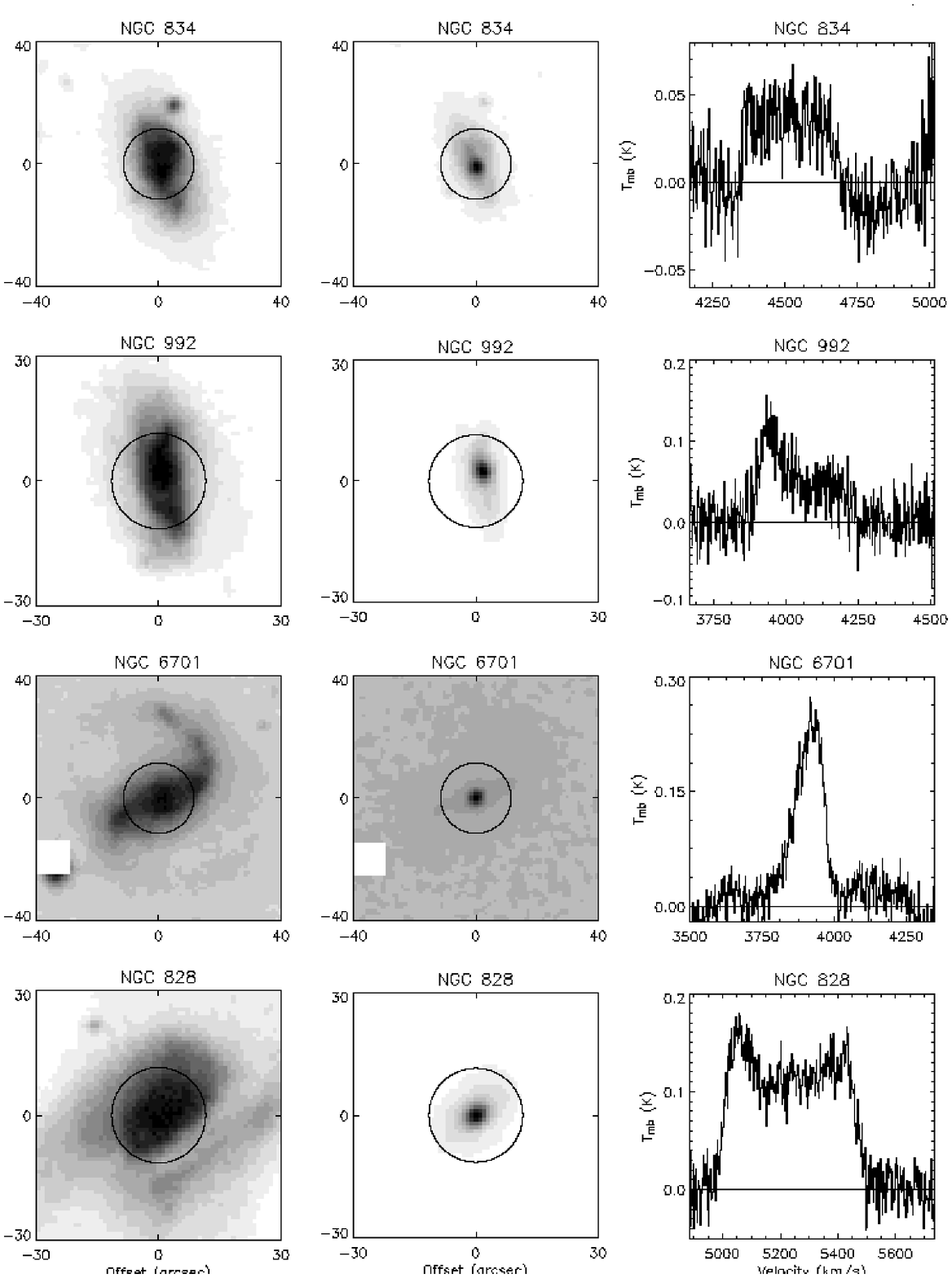}
\newpage
\plotone{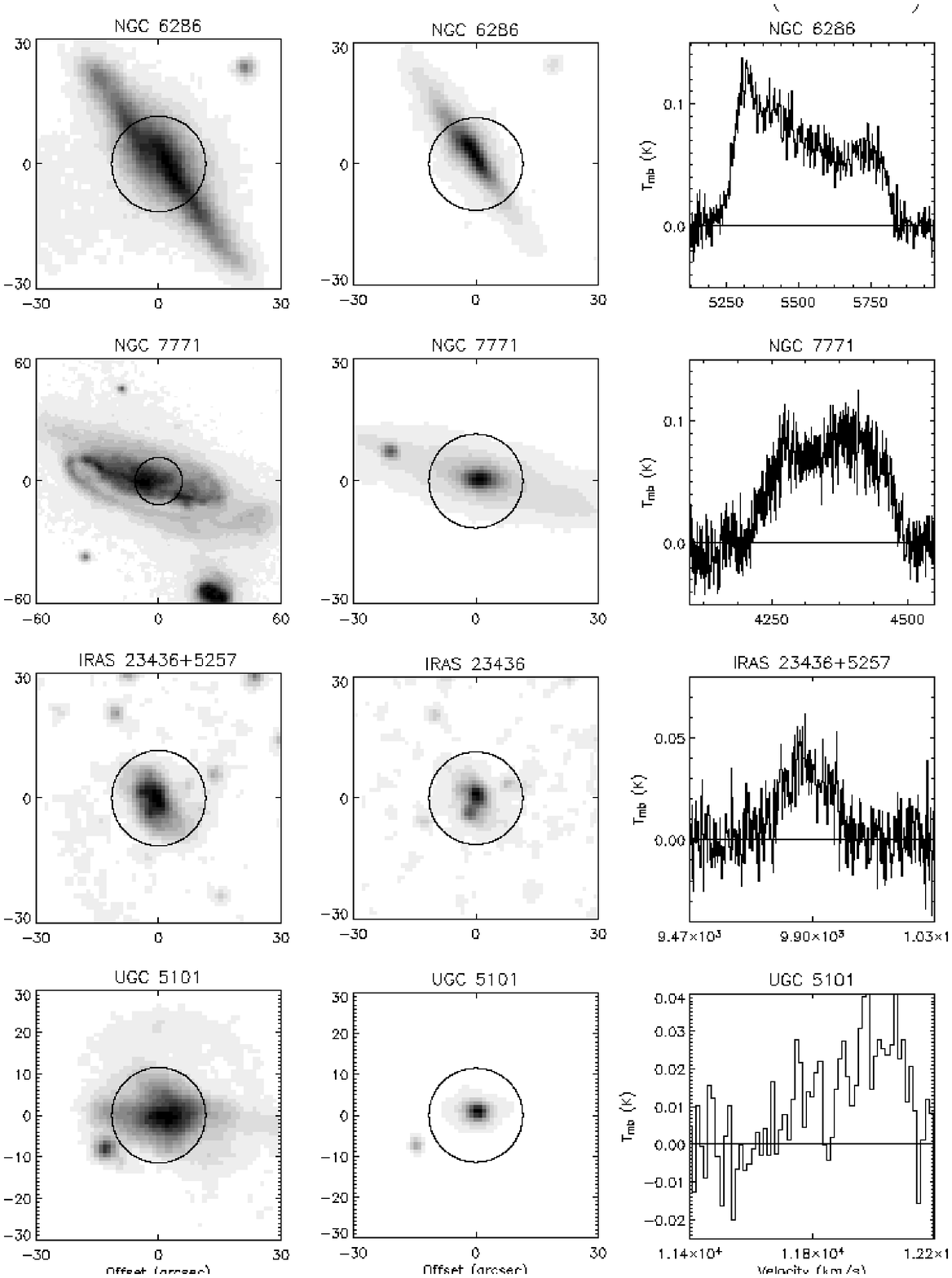}
\newpage
\plotone{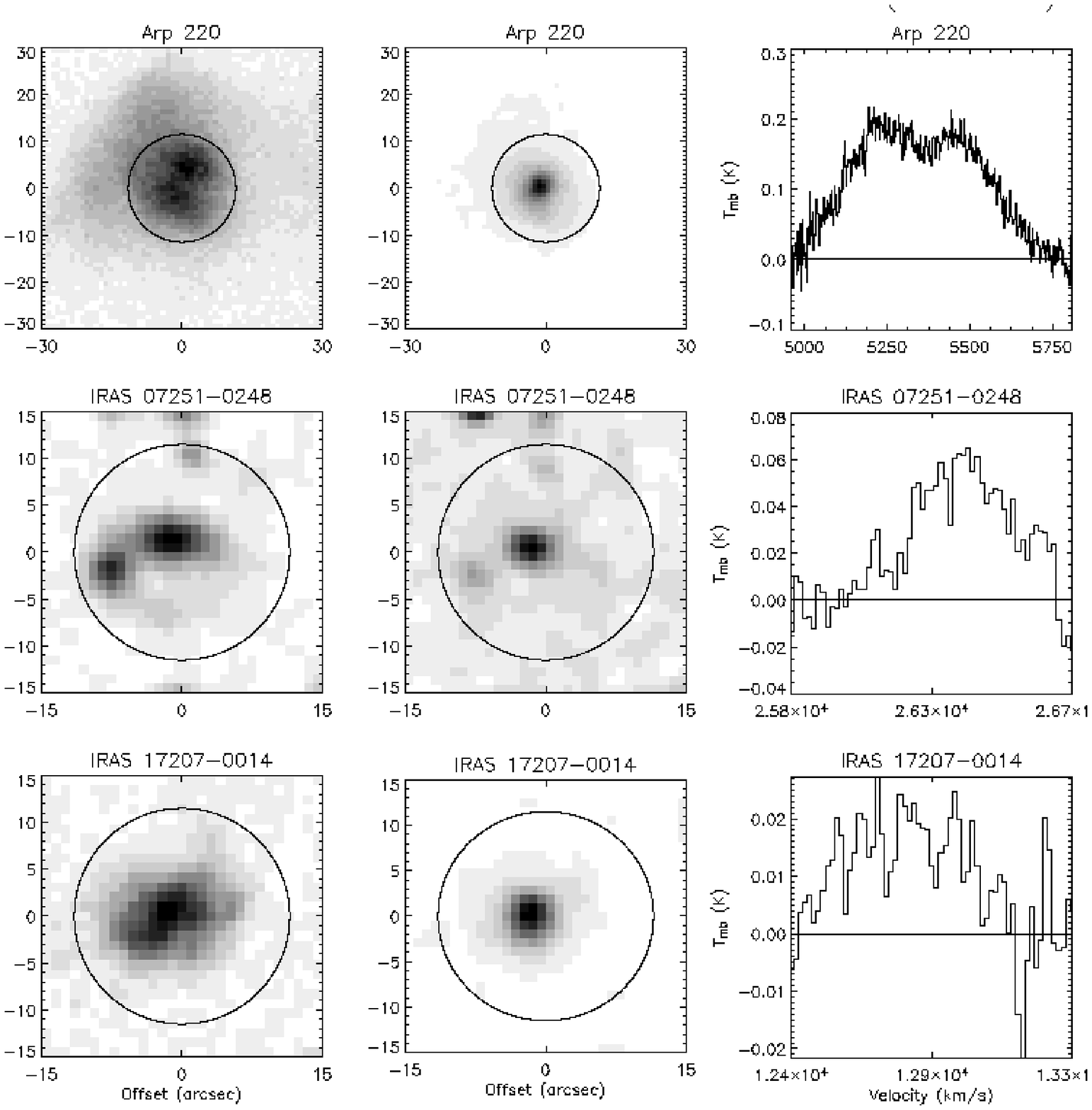}
\newpage
\plotone{f2.eps}


\newpage



\begin{thebibliography}{}
\bibitem[Armus et al. 2004]{Arm04}Armus, L., 2004, \apjs, 154, 178
\bibitem[Ashby, Houck, \& Matthews]{AHM95}Ashby, M.L.N., Houck, J.R., 
Matthews, K., 1995, \apj, 447,545
\bibitem[Barnes \& Hernquist, 1991]{BH91}Barnes, J.E., Hernquist, L.E., 
1991, \apj, 370, L65
\bibitem[Bryant \& Scoville 1999]{BaS99}Bryant, P.M., Scoville, N.Z.,
1999, \apj,117,2632
\bibitem[Carilli et al. 2004, in press]{Car04}Carilli, C.L., Solomon, P., 
Vanden Bout, P., Walter, F., Beelen, A., Cox, P., Bertoldi, F.,
Menten, K.M., Isaak, K.G., Chandler, C.J., Omont, A., 2004, \apj, 
(astro-ph/0409054)
\bibitem[Chini, Krugel, \&Lemke, 1996]{CKL96}Chini, R., Krugel, E., 
Lemke, R., 1996, A\&AS, 118, 47
\bibitem[Close et al. 1995]{Clos05}Close, L.M., Hall, P.B., 
Liu, C.T., Hege, E.K., 1995, \apj, 452L,9
\bibitem[Evans et al. 1998]{Ev98}Evans, A.S., Kim, D.C., Mazzarella, J.M., 
Scoville, N.Z., Sanders, D.B., 1999, \apj, 521,107
\bibitem[Farrah, et al. 2003]{Fa03}Farrah, D., Afonso, J., Efstathiou, A.,
Rowan-Robinson, M., Fox, M., Clements, D., 2003, \mnras, 343,585
\bibitem[Lawrence et al. 1991]{Law91}Lawrence, A., Rowan-Robinson, M.,
Leech, K., Jones, D.H.P., Wall, J.V., 1991, \mnras,240,329
\bibitem[Leech et al. 1994]{Lee94}Leech, K.J., Rowan-Robinson, M.,
Lawrence, A., Hughes, J.D., 1994, \mnras,267,253
\bibitem[Farrah et al. 2002]{Fa02}Farrah, D., Serjeant, S., 
Efstathiou, A., Rowan-Robinson, M., Verma, A., 2002, \mnras,
335,1163
\bibitem[Gao \& Solomon 2004a]{GS04b}Gao, Y., Solomon, P.M., 2004a, 
 \apj, 606,271
\bibitem[Gao \& Solomon 2004b]{GS04a}Gao, Y., Solomon, P.M., 2004b, 
\apj, 152:63.
\bibitem[Hattori et al. 2004]{Hat04}Hattori, T., Yoshida, M., 
Ohtani, H., Sugai, H., Ishigaki, T., Sasaki,  HOUCK, J. RM., Hayashi, T., 
Ozaki, S., Ishii, M., Kawai, A., 2004, AJ, 127,736
\bibitem[Imanishi 2000]{Im00}Imanishi, M., 2000, \mnras, 319,331
\bibitem[Imanishi, Dudley \& Maloney, 2001]{IDM01}Imanishi M., 
Dudley, C.C., Maloney, P.R., 2001, \apj, 558L, 93.
\bibitem[Imanishi et al. 2003]{Im03}Imanishi, M., Terashima, Y., 
Anabuki, N., Nakagawa, T., 2003, \apj, 596L, 167
\bibitem[Iono et al. 2004]{Ion04}Iono, D., Ho, P.T.P., Yun, M., 
Matsushita, S., Peck, A., Sakamoto, K., 2004, \apj, 616L,63
\bibitem[Kennicutt, 1998]{Ken98}Kennicutt,R.C., 1998, ARA\&A,36,189
\bibitem[Kormendy \& Richstone 1995]{KR95}Kormendy, J., Richstone, D., 1995, 
  ARA\&A,33,581
\bibitem[Krugel et al. 1990]{Kru90}Krugel, E., Steppe, H., Chini, R.,
1990, A\&A, 229,17
\bibitem[Maloney, P. 1989]{Mal89}Maloney, P., 1989, {\it The Interstellar 
Medium in External Galaxies - Proc. of the Second Wyoming Conf. on the 
Interstellar Medium}
\bibitem[Mauersberger et al. 1999]{Mau99}Mauersberger, B., Henkel, C.,
Walsh, W., Schulz, A., 1999, A\&A,341,256
\bibitem[Mihos \& Hernquist, 1996]{MH96}Mihos, J.C., Hernquist, L.E., 
1996, \apj, 464,641
\bibitem[Mirabel et al. 1990]{Mir90}Mirabel, I.F., Booth, R.S., Garay, G., 
Johansson, L.E.B., Sanders, D.B., 1990, A\&A236,327
\bibitem[Mooney \& Solomon 1988]{MS98}Mooney, T.J., Solomon, P.M., 1988, 
\apj, 334, L51
\bibitem[Rigopoulou et al. 1996]{Rig96} Rigopoulou, D., Lawrence, A.,
White, G.  J., Rowan-Robinson, M., Church, S.E., 1996, A\&A, 305,747
\bibitem[Sanders et al. 1988]{San88}Sanders, D.B., Soifer, B.T.,
Elias, J.H., Madore, B.F., Matthews, I., Neugebauer, G., Scoville,
N.Z., 1988, \apj, 325, 74S
\bibitem[Sanders, Scoville \& Soifer 1991]{SSS91}  Sanders, D.B.,
Scoville, N.Z., Soifer, B.T., 1991, \apj, 370,158 [SSS91]
\bibitem[Sanders et al. 1993]{San93}Sanders, D.B., in Back to the 
Galaxy, ed. F. Verter (Dordrecht: Kluwer), 1993, 311
\bibitem[Sanders \& Mirabel, 1996]{SaM96}Sanders, D.B., Mirabel, I.F.,
1996, \araa, 34,749
\bibitem[Sanders et al. 2003]{Sa03}Sanders, D.B., Mazzarella, J.M., 
  Kim, D-C., Surace, J.A., Soifer, B.T., 2003, \aj, 126:1607
\bibitem[Sakamoto et al. 1999]{Sak99}Sakamoto, K, Scoville, N.Z., Yun,
M., Crosas, M., Genzel, R., Tacconi, L.J.,1999, \apj, 514,68
\bibitem[Scoville \& Sanders, 1987]{SS87}Scoville, N.Z., Sanders, D., 
1987, in {\it Interstellar Processes}, ed. H. Thronson \& 
D. Hollenbach (Dordrecht: Reidel),21
\bibitem[Scoville et al. 1989]{Sco89}Scoville, N.Z., Sanders, D.B., 
Sargent, A.I., Soifer, B.T., Tinney, C.G., 1989, \apj, 345,25
\bibitem[Scoville et al. 1995]{Sco95}Scoville, N.Z., Yun, M.S., 
Brown, R.L., Vanden Bout, P.A., 1995, \apj, 449,L109
\bibitem[Scoville, Yun \& Bryant, 1997]{SYB97}Scoville, N.Z., Yun,
M.S., Bryant, P.M., 1997, \apj, 484,702
\bibitem[Scoville et al. 2000]{Sco00}Scoville, N.Z., Evans, A.S., 
Thompson, R., Rieke, M., Hines, D.C., Low, F.J., Dinshaw, N., 
Surace, J.A., Armus, L., 2000, AJ, 119,991S
\bibitem[Scoville, 2003]{Sco03}Scoville, N.Z., 2003, JKAS, 36:167
\bibitem[Solomon, Downes \& Radford 1992]{SDR92}Solomon, P.M., 
Downes, D., Radford, S.J.E., 1992, \apj, 387L, 55S
\bibitem[Solomon et al. 1997]{Sol97}Solomon, P.M., Downes, D., 
Radford, S.J.E., Barrett, J.W., 1997, \apj, 478,144
\bibitem[Soifer et al. 1987]{Soi87}Soifer, B.T., Sanders, D.B.,
Madore, B.F., Neugebauer, G., Danielson, G.E., Elias, J.H., Lonsdale,
C.J., Rice, W.L., 1987, \apj, 320,238
\bibitem[Soifer et al. 1989]{Soi89} Soifer, B.T., Boehmer, L.,
Neugebauer, G. Sanders, D.B., 1989, \aj, 98,3
\bibitem[Springel, Di Matteo \& Hernquist 2004a]{SDHa}Springel, V., 
Di Matteo, T., Hernquist, L., astro-ph/0409436
\bibitem[Springel, Di Matteo \& Hernquist 2004b]{SDHb}Springel, V.,
Di Matteo, T., Hernquist, L., astro-ph/0411108
\bibitem[Surace et al. 1998]{Sur98}Surace, J., Sanders, D.B., 
Vacca, W., Veilleux, S., Mazzarella, J., 1998, \apj, 492,116
\bibitem[Taniguchi \& Shioya 1998]{TS98}Taniguchi, Y., Shioya, Y.,
1998, \apjl, 501,L167
\bibitem[Tran et al. 2001]{Tr01}Tran, Q.D., Lutz, D., Genzel, R., 
Rigopoulou, D., Spoon, H.W.W., Sturm, E., Gerin, M., Hines, D.C.,
Moorwood, A.F.M., Sanders, D.B., Scoville, N., Taniguchi, Y., 
Ward, M., 2001, \apj, 552,527
\bibitem[Tinney et al. 1990]{Tin90}Tinney C.G., Scoville, N.Z.,
Sanders, D.B., Soifer, B.T.,1990, \apj, 362,473
\bibitem[Veilleux, Kim \& Sanders, 2002]{VKS02}Veilleux, S., Kim, D.-C., 
Sanders, D.B., 2002, \apjs, 143,315
\bibitem[Walter et al. 2004]{Wal04}Walter, F., Carilli, C., 
Bertoldi, F., Menten, K., Cox, P., Lo, K.Y., Fan, X., Strauss, M., 2004, 
\apj, 615,L17
\bibitem[Wang et al. 2001]{Wan01}Wang, Z., Scoville, N.Z., 
Sanders, D.B., 1991, \apj, 368,112
\bibitem[Wang et al. 2004]{Wan04}Wang, J., Zhang, Q., Wang, Z., 
Ho, P.T.P., Fazio, G., Wu, Y., 2004, \apj, 616L, 67
\bibitem[Yao et al. 2003]{Yao03} Yao, L., Seaquist, E.R., Kuno, N.,
Dunne, L., 2003, \apj, 588,771
\bibitem[Young \& Scoville 1982]{YS82}Young, J., Scoville, N., 1982, \apj,
258,467
\bibitem[Young et al. 1995]{Y85}Young, J. et al. 1995, \apjs, 98,219
\bibitem[Yun \& Scoville, 1998]{YS98}Yun, M., Scoville, N.Z., 1998,
\apj, 507,774
\end{thebibliography}
\end{document}